\documentclass[preprint,showpacs,preprintnumbers,amsmath,amssymb]{revtex4}
\usepackage{graphicx}% Include figure files

\begin{document}

\title{New Physics with IceCube}

\author{Matias M. Reynoso}
\author{Oscar A. Sampayo}
\affiliation{Departamento de F\'{\i}sica,
Universidad Nacional de Mar del Plata \\
Funes 3350, (7600) Mar del Plata, Argentina}

\email{sampayo@mdp.edu.ar}

\begin{abstract}
IceCube, a cubic kilometer neutrino telescope will be capable of
probing neutrino-nucleon interactions in the ultrahigh energy
regime, far beyond the energies reached by colliders. In this
article we introduce a new observable that combines several
advantages: it only makes use of the upward-going neutrino flux, so
that the Earth filters the atmospheric muons, and it is only weakly
dependent on the initial astrophysical flux uncertainties.
\end{abstract}


\maketitle

\section{Introduction}

High precision or high energy are necessary to study the matter at
short distances. The most powerful accelerators are the cosmic ones
in the outer space. We know that the Earth is hit by cosmic rays of
very high energy, which means that there have to be astrophysical
mechanisms capable of accelerating protons to those high energies.
It is then also possible that the same mechanisms could produce
neutrinos of high energy or that the protons and radiation produced
interact with matter to originate extremely high energy neutrinos.
Some candidate neutrino sources are Active Galactic Nuclei (AGN)
\cite{manh}, which are the central regions of certain galaxies where
the radiation emitted is comparable to the total radiation from the
entire galaxy, and Gamma Ray Bursts (GRB) \cite{waxman}, that are
the most powerful explosions since the Big Bang resulting usually
from the core collapse of massive stars.

The integrated effect over all astrophysical sources in the sky
where such producing mechanisms may operate, is expected to lead
to a diffuse neutrino flux that could be detected by IceCube
\cite{icecu}. This next-generation neutrino telescope is planned
to have a good directional resolution, a fact that will be useful
for our purposes. Another source of neutrinos that contributes to
a diffuse flux is given by the collisions of cosmic rays with the
nucleons of the atmosphere, however for energies above $10^5$ GeV,
the extraterrestrial diffuse flux should start to dominate over
the atmospheric spectrum.

These high energy neutrinos coming from different sources can be
used to look for new physics effects in neutrino-nucleon
interactions using the nucleons of the Earth as targets. In order to
bound such effects, the different observables that have been
studied, basically arise from comparing the upward-going flux that
survives after passing through the Earth (which is strongly
dependent on the neutrino-nucleon cross section) to the standard
model prediction \cite{ralston,lepto,stasto,carimalo}.

\section{Observing new physics}

In this work we define the observable $\alpha(E)$ in the following
way. Considering only upward-going neutrinos, that is, the ones with
arrival directions $\theta$ such that $0<\theta<\pi/2$, we denote by
$\alpha$ the angle such that the number of events for
$0<\theta<\alpha$ equals the number of events for
$\alpha<\theta<\pi/2$. Clearly, the value of $\alpha$ is energy
dependent. At low energies the cross section is lower and the Earth
is essentially transparent to neutrinos. In this case,
$\alpha\rightarrow \pi/3$ since for a diffuse isotropic flux this
angle divides the hemisphere into two sectors with the same solid
angle. Obviously, for extremely high energies where most neutrinos
are absorbed, $\alpha\rightarrow \pi/2$, and for intermediate
energies $\alpha$ varies accordingly between these two limiting
behaviors.

We will consider only the diffuse neutrino flux from
extraterrestrial origin and assume that it is isotropic. The use of
the observable $\alpha(E)$ reduces the effects of the experimental
systematics and initial flux dependence. The functional form of
$\alpha(E)$ sharply depends on the interaction cross section
neutrino-nucleon. In this conditions, if physics beyond the standard
model operates at these high energies, it will become manifest
directly on the function $\alpha(E)$.

If we take into account a diffuse neutrino flux and for simplicity
we consider it as decreasing with energy (an example is the given by
the Waxman and Bacall \cite{waxman-bahcall}), then in a first
analysis we can safely neglect the regeneration effects and
approximate the surviving neutrino flux by \cite{nicolaidis},
\begin{equation}\label{transporte}
\Phi(E,\theta)=\phi_0(E) e^{-\sigma_{\rm tot}(E) \tau(\theta)},
\end{equation}
where $\tau(\theta)$ is the number of nucleons per unit area in the
neutrino path through the Earth,
\begin{equation}\label{tau}
\tau(\theta)=N_A \int_0^{2 R_{\rm E}\cos\theta} \rho(z) dz.
\end{equation}
Here $\Phi_0(E)$ is the initial neutrino flux, $N_A$ is the
Avogradro number, $R_{\rm E}$ is the radius of the Earth, and
$\theta$ is the nadir angle taken from the downward-going normal to
the neutrino telescope.

%Up-going muon events from CC $\nu_{\mu}$ interactions producing an
%energetic muon traversing the detector.
Upward-going neutrinos can originate, though CC $\nu_\mu N$
interactions, energetic muons that will traverse the detector. This
is the traditional observation mode, in which the background due to
atmospheric muons is eliminated. Simulations based on AMANDA data
indicate that the direction of muons can be determined to sub-degree
accuracy and their energy can be measured to better than $30\%$ in
the logarithm of the energy. The important advantage of this mode is
the angular sub-degree resolution. On the other hand, as it was
recognized in Ref. \cite{lepto}, in IceCube we will have sufficient
energy resolution to separately assign the energy fractions in the
muon track and the hadronic shower allowing the determination of the
inelasticity distribution and the neutrino energy. Recently, the
possibility to measure the inelasticity distribution was used to
study the possibility to place bounds to new effects coming from
leptoquarks or Black-Hole production over kinematic regions never
tested before \cite{lepto,bhanchor}. In our particular case, the
possibility of independently measuring the muon energy and the
hadronic shower energy will allow us to have a reasonable
$\nu_{\mu}$-energy determination. Hence, in the following we shall
take the $\nu_{\mu}$-energy bin partition interval as $\Delta
\log_{10}E=0.5$.

To a good approximation, the expected number of events at IceCube in
the energy interval $\Delta E$ and in the angular interval $\Delta
\theta$ is given by

\begin{equation}
{\mathcal N}=n_{\rm T} T \int_{\Delta\theta}\int_{\Delta E} d\theta
dE_{\nu} \sigma(E_{\nu}) \Phi(E,\theta),
\end{equation}

where $n_{\rm T}$ is the number of target nucleons in the effective
volume, $T$ is the running time, and $\sigma(E_{\nu})$ is the
neutrino-nucleon cross section. In our analysis we are interested
only in CC contained events, for which an accurate measurement of
the inelasticity can be obtained. We take as the detection volume
for contained events the instrumented volume for IceCube which is
roughly 1 km$^3$ and corresponds to $n_{\rm T}\simeq 6 \times
10^{38}$.

\begin{figure*}
\includegraphics[angle=270,totalheight=8cm,bb= 70 180 550 680]{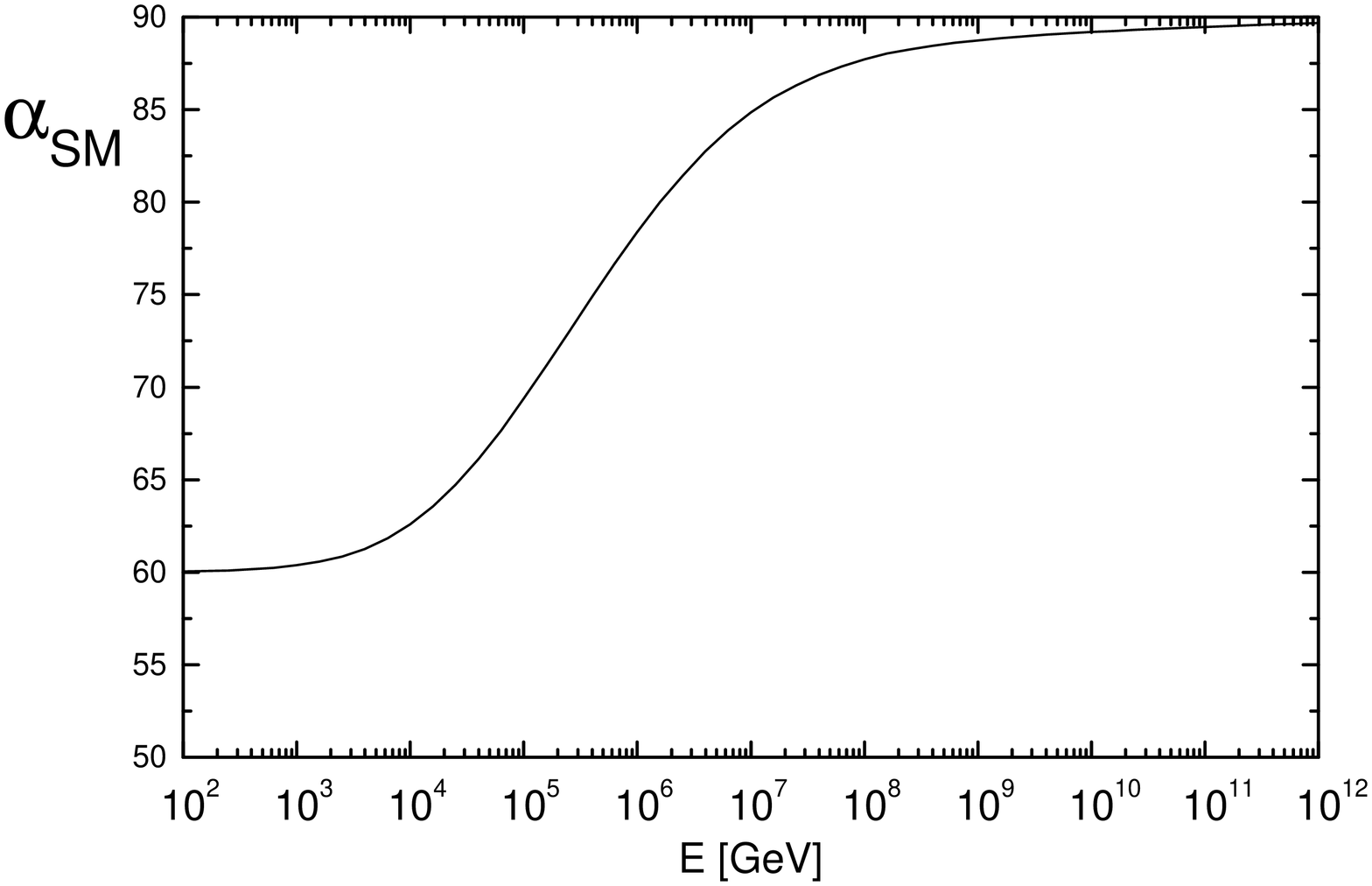}
\caption{\label{fig:alfsm} Standard Model prediction for
$\alpha_{\rm SM}(E)$}
\end{figure*}

%If we then consider energy bins of width $\Delta \log_{10}E=0.5$ and
Since the definition of $\alpha$ is the equality between two
number of events, then to a good approximation for each energy bin
all the previous factors cancel except the integrated fluxes at
each side. Thus, $\alpha$ can be defined by the equation

\begin{equation}\label{alfadef}
\int_0^{\alpha_{\rm SM}(E)}d\theta \sin\theta e^{-\sigma^{SM}(E)
\tau(\theta)}=\int_{\alpha_{\rm SM}(E)}^{\pi/2} d\theta \sin\theta
e^{-\sigma^{SM}(E) \tau(\theta)},
\end{equation}

which is numerically solved to give the results shown in
Fig.~\ref{fig:alfsm}. There we have considered the standard model
cross section as it was calculated in \cite{gandhi}, and for
$\tau(\theta)$ we use Eq.~\ref{tau} with the Earth density as given
by the PREM \cite{premm}.

Both the SM predictions for the total cross section and the Earth
density have uncertainties that propagate into the observable
$\alpha(E)$. In Fig.~\ref{fig:alfa}, where the new physics effects
on $\alpha(E)$ are shown, the effects of the mentioned uncertainties
of the observable are also included.

To reduce background, one looks for upward-moving muons produced
when neutrinos coming from the opposite side of the Earth interact
with the nucleons in their path. Directional reconstruction of these
tracks suppresses the atmospheric muons background. By selecting the
neutrino events that are upward-going it is possible to eliminate
the atmospheric muons since they are downward-going.

\section{Four-fermion interactions}

In order to model in a general way the new physics effects on
$\alpha(E)$, we consider general 4-fermion interactions as given by
an effective operator that includes also the SM fields involved in
the neutrino-nucleon scattering with left-handed neutrinos. If there
are new interactions between quarks and leptons, then the new
effects should appear at an energy enough high. We call this
characteristic energy scale for the new interactions $\Lambda$. At
energies below $\Lambda$, these interactions are suppressed by an
inverse power of $\Lambda$. Thus, the dominant effects should come
from the lowest dimensional interactions with 4-fermions
\cite{peskin},

\begin{equation}\label{lagrangeano}
\begin{split}
{\mathcal L}={\mathcal L}^{SM}+
\frac{g_N^2}{2\Lambda^2}[\eta_{LL}& ( \bar l\gamma_{\mu} P_L \nu
\medspace \bar q_j \gamma^{\mu} P_L q_i + \bar \nu \gamma_{\mu}
P_L \nu \medspace \bar q_i \gamma_{\mu} P_L q_i) +
\\ \eta_{LR}& ( \bar l\gamma_{\mu} P_L \nu \medspace \bar q_j \gamma^{\mu}
P_R q_i + \bar \nu \gamma_{\mu} P_L \nu \medspace \bar q_i
\gamma_{\mu} P_R q_i) ],
\end{split}
\end{equation}
for left-handed neutrinos, where we take $g_N^2= 4\pi$, and the
coefficients $\eta_{LL}$ and $\eta_{LR}$ can take the values $-1$,
$0$ and $1$. Choosing different values of $\Lambda$, $\eta_{LL}$,
and $\eta_{LR}$, we can test the $\alpha$ observable on different
scenarios of new physics.

\begin{figure*}
\includegraphics[angle=0,width=8cm,bb= 180 180 500 680]{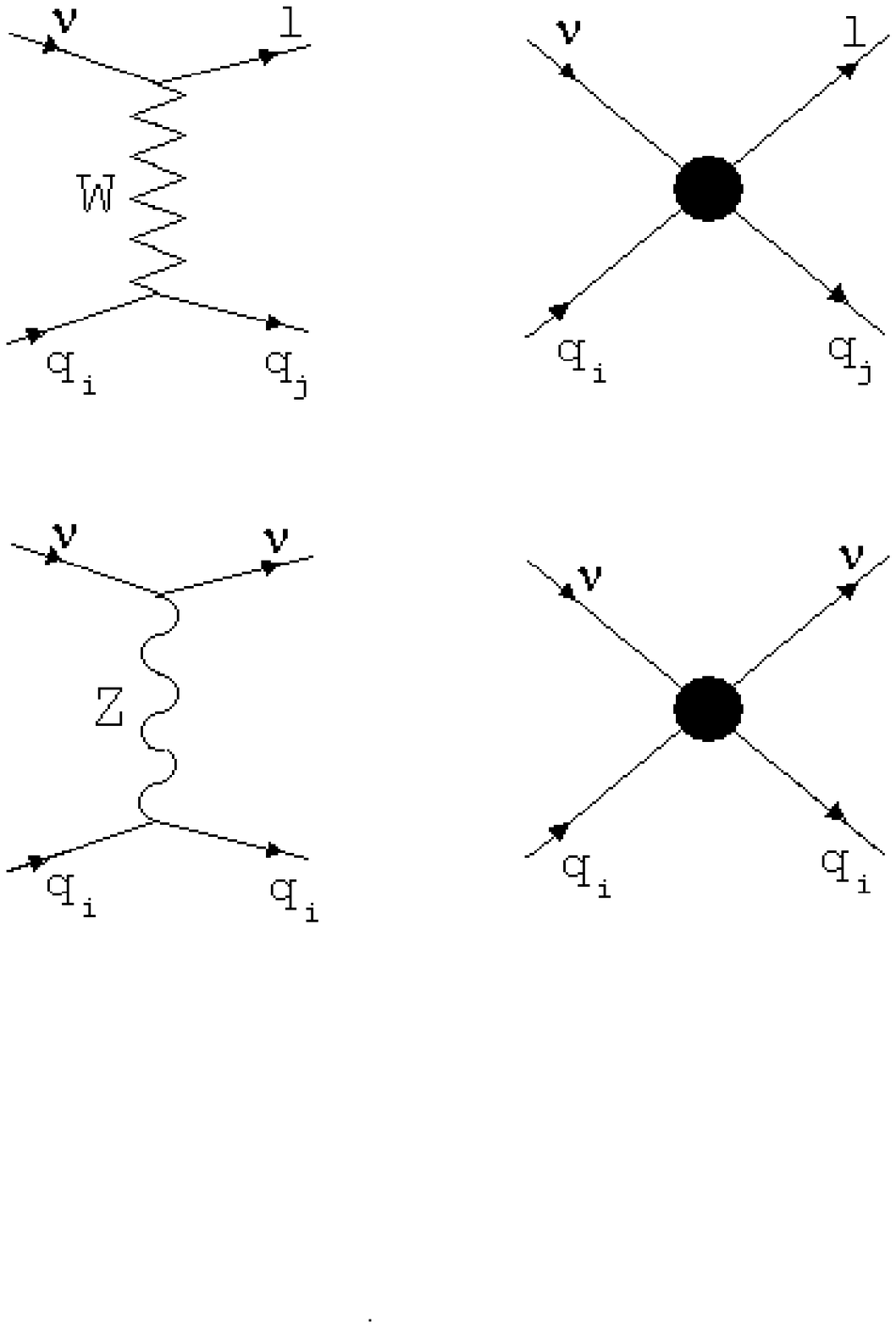}
\caption{\label{fig:diag}  Feynman diagrams contributing to charged
current and neutral current processes.}
\end{figure*}

Using the effective operator we can calculate their contribution to
the neutrino-nucleon inclusive cross section

\begin{equation}
\nu N\rightarrow \mu + \rm{anything},
\end{equation}

where $N\equiv \dfrac{n+p}{2}$ for an isoscalar nucleon. The
corresponding processes are pictured in Fig.~\ref{fig:diag} for
charged and neutral currents. The calculation is standard and we use
it to compare with the results of \cite{gandhi}. For the charged
current, the scattering amplitude is

\begin{equation}\label{amplcc}
{\cal M}^{\rm CC}=-\frac{i g^2}{2(Q^2+M_W^2)} \bar l\gamma_{\mu} P_L
\nu \medspace  \bar q_j \gamma^{\mu}(g^{'}_L P_L+g^{'}_R P_R )q_i,
\end{equation}

where
\begin{equation}
\begin{split}
g^{'}_L&
=1-\dfrac{(Q^2+M_W^2)}{\Lambda^2}\eta_{LL}\dfrac{g_{N}^2}{g^2}
\\\\
g^{'}_R&
=-\dfrac{(Q^2+M_W^2)}{\Lambda^2}\eta_{LR}\dfrac{g_{N}^2}{g^2}
\end{split}
\end{equation}
include the new physics effects.

The differential cross section for charged currents reads

\begin{equation}\label{disfsigcc}
\begin{split}
 \frac{d\sigma^{\rm CC}}{dxdy}=\frac{G_F^2 s}{\pi}\left(
\frac{M_W^2}{(Q^2+M_W^2)} \right)^2x[g^{'2}_L (Q^{\rm CC}+(1-y)^2
\bar Q^{\rm CC})   \\ +g^{'2}_R( \bar Q^{\rm CC}+(1-y)^2 Q^{\rm
CC})],
\end{split}
\end{equation}

where for an isoscalar target we have the quark distribution
functions

\begin{equation}\label{districc}
\begin{split}
Q^{\rm CC}(x,Q^2)& =\dfrac{u_v(x,Q^2)+d_v(x,Q^2)}{2}+\dfrac{u_s(x,Q^2)+d_s(x,Q^2)}{2} \\
  & + s_s(x,Q^2)+b_s(x,Q^2)
\\\\
\bar Q^{\rm CC}(x,Q^2)&
=\dfrac{u_s(x,Q^2)+d_s(x,Q^2)}{2}+c_s(x,Q^2)+t_s(x,Q^2)
\end{split}
\end{equation}

Similarly, the neutral current amplitude is

\begin{equation}\label{amplnc}
{\cal M}^{\rm NC}=-\frac{i g^2}{2 c_W (Q^2+M_Z^2)} \bar \nu
\gamma_{\mu} P_L \nu \medspace  \bar q_i \gamma^{\mu}(g^{'}_L
P_L+g^{'}_R P_R )q_i,
\end{equation}
where $c_W=\cos \theta_W$ and

\begin{equation}\label{cttnc}
\begin{split}
g^{'i}_L=g^i_L-\frac{Q^2+M_Z^2}{\Lambda^2} c_W^2 \eta_{LL}
\frac{g_{LL}^2}{g^2}  \\
g^{'i}_R=g^i_R-\frac{Q^2+M_Z^2}{\Lambda^2} c_W^2 \eta_{LR}
\frac{g_{LR}^2}{g^2}
\end{split}
\end{equation}
include the new physics effects. Here, $g^U_L=1/2-2/3 x_W$,
$g^D_L=-1/2+1/3 x_w$, $g^U_R=-2/3 x_W$, and $g^D_R=1/3 x_W$.

The neutral current differential cross section is then

\begin{equation}\label{disfsignc}
\begin{split}
\frac{d\sigma^{\rm NC}}{dxdy}=\frac{G_F^2 s}{\pi}\left(
\frac{M_Z^2}{(Q^2+M_Z^2)} \right)^2 \sum_{i=U,D}x[g^{`i2}_L
(Q^{i}+(1-y)^2 \bar Q^{i})  \\ +g^{`i2}_R( \bar Q^{i}+(1-y)^2
Q^{i})],
\end{split}
\end{equation}

where the corresponding parton distributions for a isoscalar target
read

\begin{equation}\label{partondistnc}
\begin{split}
Q^U(x,Q^2)=\dfrac{u_v(x,Q^2)+d_v(x,Q^2)}{2}+\dfrac{u_s(x,Q^2)
+d_s(x,Q^2)}{2}  \\+c_s(x,Q^2)+t_s(x,Q^2)
 \\
Q^D(x,Q^2)=\dfrac{u_v(x,Q^2)+d_v(x,Q^2)}{2}+\dfrac{u_s(x,Q^2)
 +d_s(x,Q^2)}{2} \nonumber \\+s_s(x,Q^2)+b_s(x,Q^2)
  \\ \bar
Q^U(x,Q^2)=\dfrac{u_s(x,Q^2)+d_s(x,Q^2)}{2}+c_s(x,Q^2)+t_s(x,Q^2)
 \\ \bar
Q^D(x,Q^2)=\dfrac{u_s(x,Q^2)+d_s(x,Q^2)}{2}+s_s(x,Q^2)+b_s(x,Q^2).
\end{split}
\end{equation}

\section{Results}

In order to evaluate the impact of the observable $\alpha$ to bound
new physics effects, we have estimated the corresponding
uncertainties. For the number of events we have considered it as
distributed according to a Poisson distribution and we have
propagated onto the angle $\alpha$. The number of events $N$ as a
function of $\alpha_{\rm SM}$  is
\begin{equation}
N=2 \pi n_{\rm T} T \Delta E \sigma(E) \phi_0(E) \int_0^{\alpha_{\rm
SM}}d\theta \sin\theta e^{-\sigma_T(E)\tau(\theta)}
\end{equation}

where we have considered the effective volume for contained events
for which a accurate and simultaneous determination of the muon
energy and shower energy is possible. For IceCube it corresponds to
the instrumented volume, roughly 1 km$^3$, that implies a number of
target nucleons $n_{\rm T}\simeq 6 \times 10^{38}$. For $T$ we have
taken an integration time of $15$ yr corresponding to the lifetime
of the experiment.

%If we propagate the error for $N$, the error with which $\alpha$ is
%obtained into the one in
To propagate the error on $N$ to obtain the one on $\alpha$, we note
that
\begin{equation}
\Delta N=\dfrac{dN}{d\alpha} \Delta\alpha,
\end{equation}

and dividing by $N$ we obtain for $\Delta\alpha$,

\begin{equation}
\Delta\alpha=\left[\int^{\alpha_{\rm SM}(E)}_0 d\theta \left(
\dfrac{\sin\theta}{\sin\alpha_{\rm SM}(E)} \right)
e^{\sigma_T(E)[\tau(\alpha_{\rm SM}(E))-\tau(\theta)]}\right] \left(
\dfrac{\Delta N}{N} \right),
\end{equation}

where for Poisson distributed events we have

\begin{equation}
\Delta N=\sqrt{N}.
\end{equation}

In the evaluation of the errors on $\alpha$ it is necessary to
consider the initial flux $\phi_0^{\nu_{\mu}}$ to estimate the
events rates.

The usual benchmark here is the so-called Waxman-Bahcall (WB) flux
for each flavor,

\begin{equation}
E_{\nu_{\mu}}^2 \phi_{WB}^{\nu_{\mu}}\simeq 2\times 10^{-8} {\rm
GeV} \ {\rm cm}^{-2} {\rm s}^{-1} {\rm sr} ^{-1},
\end{equation}

which is derived assuming that neutrinos come from transparent
cosmic ray sources \cite{waxman-bahcall}, and that there is adequate
transfer of energy to pions following $pp$ collisions. However, one
should keep in mind that if there are in fact hidden sources which
are opaque to ultra-high energy cosmic rays, then the expected
neutrino flux will be higher.

On the other hand, we have the experimental bound set by AMANDA. A
summary of the bounds can be found in Refs.
\cite{desiati,amanda-bound} and a representative value for it is

\begin{equation}
E_{\nu_{\mu}}^2 \phi_{AM}^{\nu_{\mu}}\simeq  2 \times 10^{-7} {\rm
GeV} \ {\rm cm}^{-2} {\rm s}^{-1} {\rm sr} ^{-1}.
\end{equation}

However, with the intention to estimate the number of events we have
considered an intermediate flux level slightly below the present
AMANDA experimental bound,

\begin{equation}
E_{\nu_{\mu}}^2 \phi_{IN}^{\nu_{\mu}}\simeq 10^{-7} {\rm GeV} \ {\rm
cm}^{-2} {\rm s}^{-1} {\rm sr} ^{-1},
\end{equation}

which is the flux that we have used to estimate the uncertainties on
the angle $\alpha$.

\begin{figure*}
\includegraphics[angle=270,totalheight=8cm,bb= 70 180 550 680]{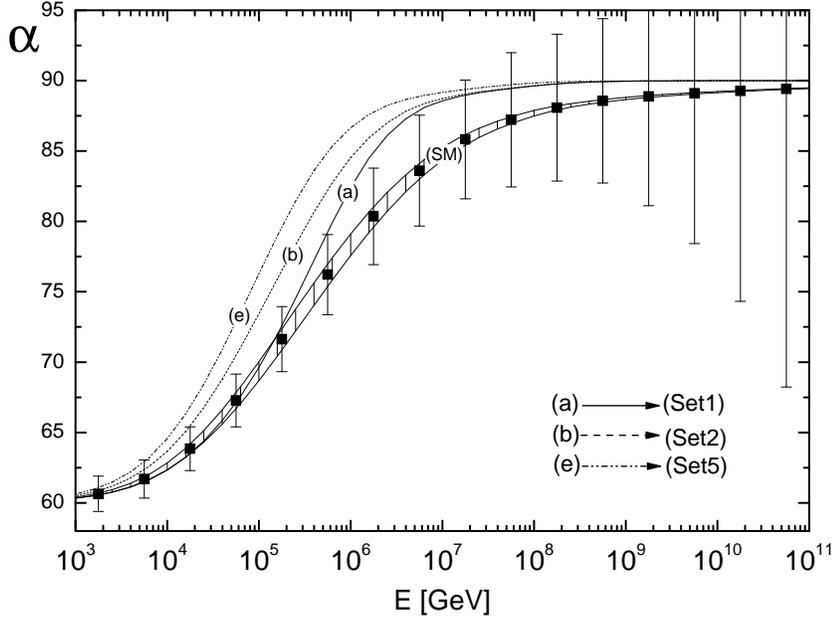}
\caption{\label{fig:alfa} $\alpha(E)$ for different sets of
parameters: (a) set1, (b) set2, (e) set5, and the standard model
prediction with the theoretical uncertainties included (Shaded
region) and the errors in $\alpha$ as resulting from the ones in the
events numbers.}
\end{figure*}

In Fig.~\ref{fig:alfa} we show our results for the $\alpha$
observable for the most representative sets of parameters and the
standard model prediction, including the theoretical uncertainties
from the SM cross section, the Earth density, and the errors coming
from the statistical uncertainties in the number of events. For the
new physics effects we have considered the sets of parameters shown
in Table~\ref{tab:table1}.

\begin{table}
\caption{\label{tab:table1}Sets of parameters for the new four-
fermion contact interactions.}
\begin{ruledtabular}
\begin{tabular}{ccccc}
 Set & $\eta_{LL}$ & $\eta_{LR}$ & $\Lambda$ (TeV) \\
\hline
 1 & 1 & 1 & $1$  \\
 2 & -1  & -1 & $1$  \\
 3 & -1  & -1 & $2$  \\
 4 & 1   &  1 & $0.8$  \\
 5 &  -1  & -1  & $0.8$  \\
\end{tabular}
\end{ruledtabular}
\end{table}

\begin{figure*}
\includegraphics[angle=270,totalheight=8cm,bb= 70 180 550 680]{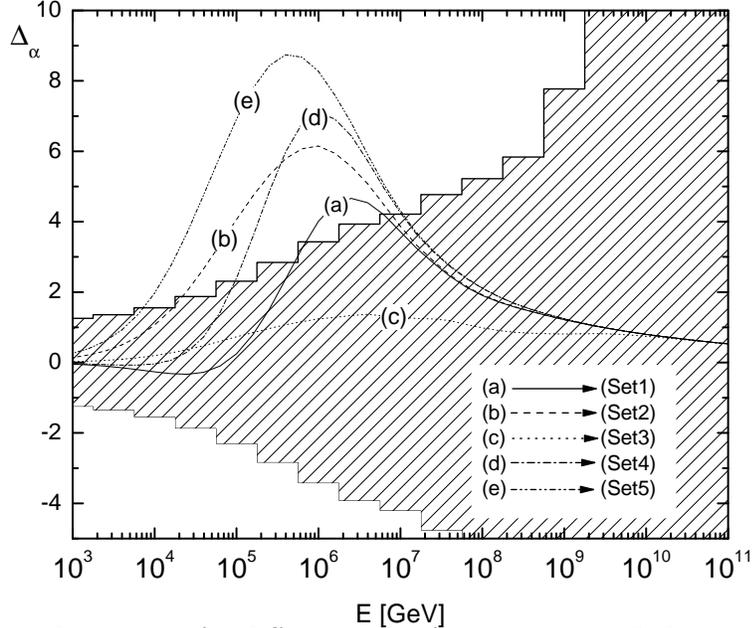}
\caption{\label{fig:dif}  Differences between $\alpha$ for different
sets of parameters and the standard model prediction
($\Delta_{\alpha}=\alpha_{\rm SM}-\alpha_{{\rm Set}_i}$). We have
included the uncertainties on $\Delta_\alpha$ calculated for the
intermediate initial flux $\phi_{IN}$ (Shaded region). The bin
partition energy interval is taken as $\Delta \log_{10}E=0.5$}
\end{figure*}

In Fig.~\ref{fig:dif} we show the differences between the values of
$\alpha$ for different sets of parameters and the standard model
values as a function of the energy. It can be seen that the maximum
sensitivity is reached in the intermediate energy range ($10^5
\rm{GeV} <E<10^7 \rm{GeV}$). In this figure we also include the
uncertainties in the Standard Model prediction coming from the ones
in the event numbers (Shaded region).

In the same context, we can define another observable related to
$\alpha_{\rm SM}(E)$. We consider the hemisphere $0<\theta<\pi/2$
divided into two regions by the angle $\alpha_{\rm SM}$, ${\cal
R}_1$ for $0<\theta<\alpha_{\rm SM}$ and ${\cal R}_2$ for
$\alpha_{\rm SM}<\theta<\pi/2$. We then calculate the ratio $\chi$
between the number of events for each region,

\begin{equation}
\chi=\frac{N_1}{N_2},
\end{equation}

where $N_1$ is the number of events in the region ${\cal R}_1$ and
$N_2$ is the number of events in the region ${\cal R}_2$. By using
$\chi$, the effects of experimental systematics and initial flux
dependence are also reduced.

\begin{figure*}
\includegraphics[angle=270,totalheight=8cm,bb= 70 180 550 680]{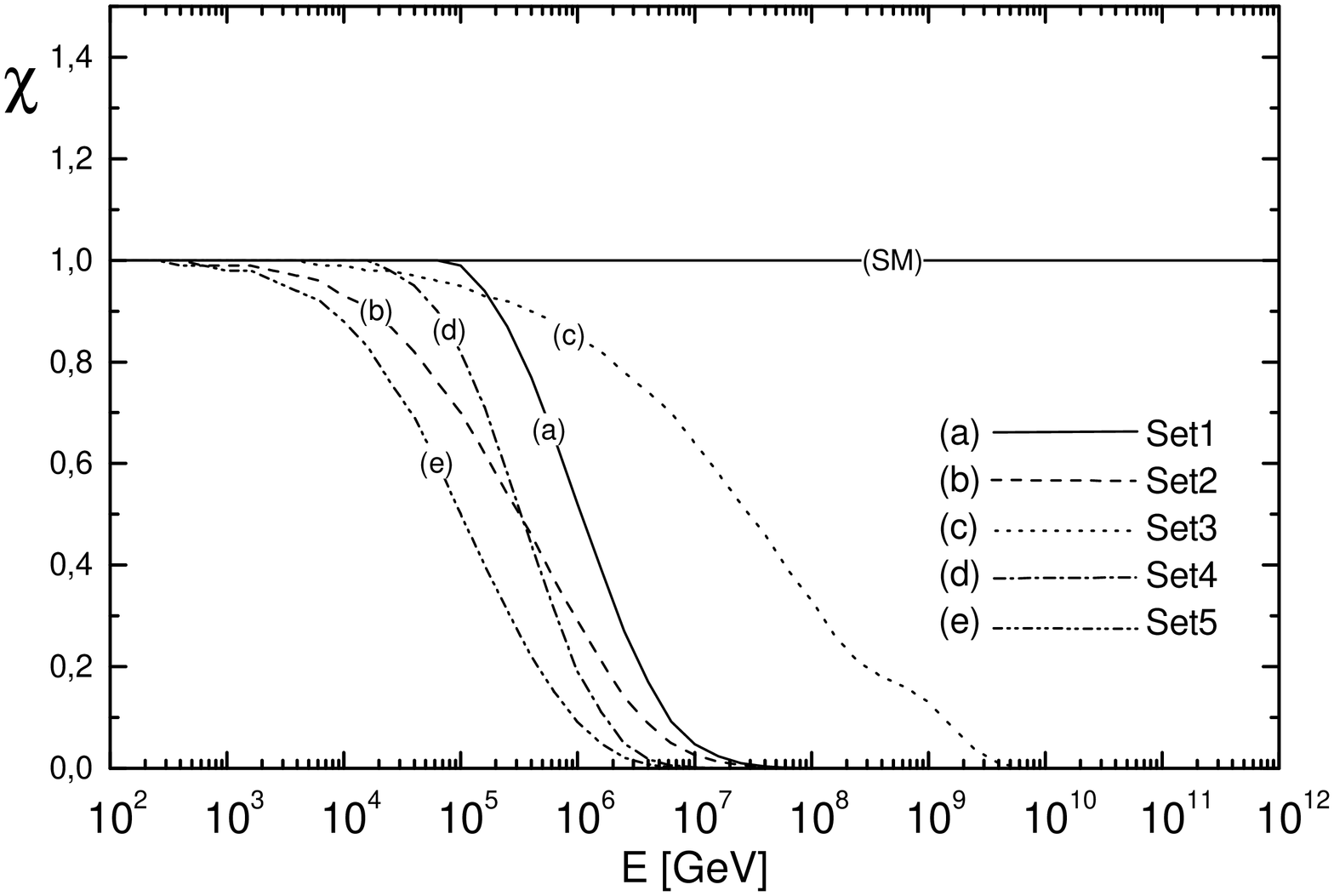}
\caption{\label{fig:capa}Observable $\chi(E)$ for different sets
of parameters.}
\end{figure*}

If there is only standard model physics, then we have that the
ratio $\chi=1$. The new physics effects produce a deviation from
the standard value as we show in Fig.~\ref{fig:capa} for the
different sets of parameters (different values of $\Lambda$,
$\eta_{LL}$, and $\eta_{LR}$).

\section{Conclusions}

In the present work we have studied a new observable that combines
the surviving neutrino flux after passing through the Earth in a way
that reduces the experimental systematics and the dependence with
the initial flux. This observable, the angle $\alpha$, divides the
upward-going hemisphere (with respect the arrival neutrino
directions) into two homo-event sectors and it is dependent, of
course, on the neutrino energy. The function $\alpha(E)$ is sharply
dependent on the neutrino-nucleon cross section,
%and this fact become it in a
which makes it a useful observable to bound new physics. In order to
test the sensitivity of $\alpha$, we have calculated the new physics
effects coming from four-fermion contact interactions. We have also
studied as another observable, the ratio between the number of
events in the regions defined by $\alpha_{\rm SM}(E)$.

We note that the introduced observable present no deviation from
the standard model prediction for low energies ($<10^3$ GeV) at
which almost no interactions occur with or without new physics. At
high energies ($>10^9$ GeV) the neutrino mean free path is so
small that the integrated surviving flux of Earth skimming
neutrinos equals the one corresponding to almost the whole
hemisphere. The corresponding increase in the cross section
implies that, given the exponential behavior of the integrated
flux, a great attenuation takes place both taking into account new
physics effects or not.

Finally, we point out that this technique can be applied to any
specific case of physics beyond the standard model, which is left
for future work.

\begin{acknowledgments}
We thank CONICET (Argentina) and Universidad Nacional de Mar del
Plata (Argentina) for their financial supports.
\end{acknowledgments}

\end{document}